\documentclass[11pt,a4paper]{article}
\usepackage{jheppub}
\usepackage{braket}
\usepackage[english]{babel}
\usepackage[utf8]{inputenc}
\usepackage{amsmath}
\usepackage{array, url}
\usepackage{amsfonts}
\usepackage{amsthm}
\usepackage{bm}
\usepackage{titlesec}
\usepackage{graphicx}
\usepackage{epstopdf}
\usepackage{tensor}
\usepackage{mathrsfs}
\DeclareGraphicsExtensions{.eps}
\usepackage{caption}
\usepackage{subcaption}
\usepackage{etoolbox}
\usepackage{tikz}
\usepackage{pgfplots}
\usepackage{enumerate} 
\usepackage{comment}

\title{{\Large \textbf{Quasi-local energy and ADM mass in pure Lovelock gravity}}}
\date{}
\author[a]{Jani Kastikainen}
\affiliation[a]{Department of Physics and Helsinki Institute of Physics, University of Helsinki\\
	Helsinki, FIN-00014, Finland}
\emailAdd{jani.kastikainen@helsinki.fi}
\abstract{We study how the standard definitions of ADM mass and Brown-York quasi-local energy generalize to pure Lovelock gravity. The quasi-local energy is renormalized using the background subtraction prescription and we consider its limit for large surfaces. We find that the large surface limit vanishes for asymptotically flat fall-off conditions except in Einstein gravity. This problem is avoided by focusing on the variation of the quasi-local energy which correctly approaches the variation of the ADM mass for large surfaces. As a result, we obtain a new simple formula for the ADM mass in pure Lovelock gravity. We apply the formula to spherically symmetric geometries verifying previous calculations in the literature. We also revisit asymptotically AdS geometries.}
\arxivnumber{1908.05522}

\begin{document}

\maketitle

\section{Introduction}

Defining a measure of energy in general relativity has a long and diverse history. Traditional definition of energy as the conserved quantity corresponding to time translation invariance fails in curved spacetimes, because such a symmetry does not exist in general. Regardless, various proposals for energy in stationary spacetimes have been put forward such as the Komar mass \cite{komar_covariant_1959} and the ADM mass \cite{arnowitt_dynamics_2008,regge_role_1974}.\footnote{Note that the Komar mass is not conserved for all physically reasonable solutions \cite{misner_gravitational_1963}. The author thanks S. Deser for pointing this out.} After the development of black hole thermodynamics, the formulation of energy is also relevant from the perspective of the black hole first law (see for example \cite{wald_black_1993,iyer_properties_1994}).

ADM mass is defined as the on-shell value of the Hamiltonian of general relativity \cite{arnowitt_dynamics_2008,regge_role_1974}. It measures the total energy content of a spacetime as measured by an observer located far away from the source. If one wants to measure the energy contained within a subset $ \mathcal{C} $ of the Cauchy slice, a new quantity is needed. Such a quasi-local energy $ M_\mathcal{B} $ was first proposed by Brown and York \cite{brown_quasilocal_1993}, and it is defined as an integral over the codimension-two surface $ \partial \mathcal{C} \equiv \mathcal{B} $ of the boundary stress-energy tensor projected onto the Cauchy slice. The result is normalized by subtracting the corresponding energy of flat space. When the surface is taken to be arbitrarily large, the energy is expected to approach the ADM mass, which is indeed the case as shown in \cite{hawking_gravitational_1996} (see also \cite{brewin_simple_2007}).

Lovelock gravity is the unique theory of gravity constructed out of the Riemann tensor that has equations of motion second order in the derivatives of the metric (see \cite{padmanabhan_lanczos-lovelock_2013} for an introduction). Its action is a linear combination of Lovelock invariants that reduce to topological invariants in their corresponding critical dimensions \cite{yale_structure_2011}. Lovelock invariants are constructed out of higher powers of the Riemann tensor and contain Ricci scalar as the special case. Pure Lovelock gravities are a subset of Lovelock gravity theories labeled by a non-negative integer $ m $ that contain a single Lovelock invariant in the action. They are very similar to Einstein gravity $ m=1 $ and extend many of its properties to higher dimensions \cite{kastor_black_2006,kastor_komar_2008,dadhich_discerning_2016}.

The asymptotic behavior of solutions of pure Lovelock gravity is not the same as in Einstein gravity, but depends on $ m $. This means that the standard ADM mass formula no longer applies, because it is finite (and thus well defined) only for asymptotic fall-off conditions of Einstein gravity. A generalization is hence needed. In \cite{kastor_mass_2011}, a generalization to Lovelock gravity was derived by generalizing the Einstein gravity derivation of \cite{regge_role_1974}. They applied the resulting mass formula to asymptotically AdS spaces and found that it is simply proportional to the mass in Einstein gravity. Specializing to pure Lovelock gravity, their result then implies that the mass is independent of $ m $ which is expected, because in AdS space, the asympotic behaviour of solutions is independent of $ m $ \cite{chakraborty_1/r_2018}. The case of asymptotically flat solutions was not analyzed in \cite{kastor_mass_2011}, which is one of the gaps filled in this paper. We find that the ADM mass in flat space is not integrable for $ m\geq 2 $, but for spherically symmetric spacetimes, the mass can be integrated and we find the explicit formula. When applied to a static black hole solution, the mass agrees with the literature \cite{banados_dimensionally_1994,cai_topological_1999,crisostomo_black_2000,cai_black_2006,kastor_komar_2008}.

Pure Lovelock gravity, and Lovelock gravity in general, suffer from the same problem as Einstein gravity that the variational principle is not well defined in the presence of boundaries \cite{gibbons_action_1977,york_role_1972}. Hence the action must be supplemented by a generalization of the Gibbons-Hawking surface term to pure Lovelock gravity. Explicit formulae for the surface terms were first presented by Myers \cite{myers_higher-derivative_1987} in the language of differential forms. In the metric formulation, the surface terms for a spacelike boundary first appeared in \cite{teitelboim_dimensionally_1987} where the Hamiltonian formalism of Lovelock gravity was established. A straightforward derivation was recently presented in \cite{chakraborty_novel_2017} which also directly applies to timelike boundaries.

Given the surface terms, it should be possible to define quasi-local energy in Lovelock gravity as anticipated in \cite{chakraborty_novel_2017}. We focus on pure Lovelock gravities and try to generalize the Einstein gravity definition presented in \cite{brown_quasilocal_1993}. The idea of \cite{brown_quasilocal_1993} is to perform a Hamiltonian decomposition of the theory in the presence of a timelike boundary and then define the energy as the on-shell value of the Hamiltonian. What one also needs is a way to regularize the resulting energy: the flat space value should be zero and the large surface limit should coincide with the ADM mass of pure Lovelock gravity.

In Einstein gravity, the background subtraction method produces a finite energy with these properties \cite{hawking_gravitational_1996}. In pure Lovelock gravity $ m \geq 2 $, we find that the background subtraction fails: the large surface limit vanishes for asymptotically flat geometries. However, if one only considers variations of the energy, then the limit is finite and coincides with the variation of the ADM mass. This is the main result of our paper and it generalizes the corresponding statements proven before in Einstein gravity \cite{hawking_gravitational_1996,brewin_simple_2007}. In addition, we show that just like in Einstein gravity \cite{brown_quasilocal_1993}, the resulting quasi-local energy can be defined in terms of the boundary stress-energy tensor. These results further extend the list of similarities between Einstein gravity and pure Lovelock gravity.

Instead of using background subtraction, there are also other ways to renormalize the quasi-local energy: one example is the use of counterterms \cite{kraus_gravitational_1999,miskovic_counterterms_2007}. Another method was presented in \cite{chakraborty_brown-york_2015} where it was proven that the large surface limit of the resulting renormalized energy is proportional to the ADM mass in spherically symmetric geometries. However, the exact prefactor of proportionality was not specified. Using the explicit formula for the ADM mass, we verify their result and determine the exact prefactor. We also generalize the calculation and show that the proportionality does not extend to arbitrary geometries.

\subsection{Summary of main results}

We generalize the Einstein gravity derivations of ADM mass \cite{regge_role_1974} and Brown-York quasi-local energy \cite{brown_quasilocal_1993} to pure Lovelock gravity. The ADM mass in pure Lovelock gravity is non-integrable and given by
\begin{equation}
	\delta M_{(m)}^\text{ADM} = -\frac{1}{16\pi}\int_\infty\sqrt{\sigma}\, 2\bar{P}^{ab}_{cd(m)}n_aD^c \delta \widetilde{\gamma}^d_b
	\label{sumadm}
\end{equation}
where $ \widetilde{\gamma}_{ab} $ is the asymptotic correction to the flat metric. This is finite given asymptotically flat fall-off conditions specialized to pure Lovelock gravity. For asymptotically AdS solutions, the mass is integrable and proportional to the Einstein value $ M_{(1)}^\text{ADM} $ for all values of $ m $.

The main result of the paper is a perturbative formula for the quasi-local energy in pure Lovelock gravity:
\begin{equation}
	\delta M^{(m)}_{\mathcal{B}} =  \frac{m}{4\pi}\int_\mathcal{B}\sqrt{\sigma}\, \widehat{E}^i_{j(m-1)}\delta \widehat{K}_{i}^j.
	\label{sumquas}
\end{equation}
The non-perturbative version is not well-defined as it has a vanishing limit for large surfaces. We show that the limit of \eqref{sumquas} for large surfaces coincides with \eqref{sumadm}:
\begin{equation}
	\delta M_{(m)}^\text{ADM} = \lim_{\mathcal{B}\rightarrow \infty}\frac{m}{4\pi}\int_\mathcal{B}\sqrt{\sigma}\, \widehat{E}^i_{j(m-1)}\delta \widehat{K}_{i}^j.
\end{equation}
This formula provides a simple method for calculating the mass of a given solution, which we demonstrate explicitly in section \ref{expsph}.

The paper is structured as follows. In section \ref{puresurfs} we introduce pure Lovelock gravity and derive the generalization of the ADM mass. Next in section \ref{sectquas}, we review the generalization of the Gibbons-Hawking surface term and define the  quasi-local energy which is renormalized using the background subtraction prescription. The derived formulas are then applied to spherically symmetric geometries in section \ref{expsph}. Up to this point only asymptotically flat geometries were considered, leaving the analysis of asymptotically AdS geometries to section \ref{adscase}.

\subsection{Notation conventions}

We collect our notation here for convenience. We work with a $ D $-dimensional spacetime $ \mathcal{M} $ with the metric $ g_{ab} $ and indices $ a,b,\ldots = 0,1,\ldots, D-1 $. The spacelike Cauchy slice $ \Sigma $ has the metric $ \gamma_{ab} $ and the timelike normal $ u^a $. The spatial metric satisfies $ \gamma_{ab}u^b = 0 $ and can be written as $ \gamma_{ab} = g_{ab} + u_au_b $. Spatial indices are denoted by $ \alpha,\beta = 1,2, \ldots, D-1 $. If $ \mathcal{M} $ is taken to have a timelike boundary, it is denoted by $ \Gamma $ and it has the metric $ h_{ab} = g_{ab} - n_an_b $ with the spacelike normal $ n^a $. The indices of $ \Gamma $ are denoted by $ \mu,\nu,\ldots = 0,1,\ldots, D - 2 $. We also denote $ \mathcal{B} = \Gamma \cap \Sigma $ which has the metric $ \sigma_{ab} =  g_{ab} + u_au_b - n_an_b $ and its indices are denoted  $ i,j,\ldots = 1,2,\ldots, D-2 $. The covariant derivative of $ \mathcal{M} $ is denoted by $ \nabla_a $ and of $ \mathcal{B} $ by $ D_a $.

For generalized Kronecker deltas we assign specific notation. The spacetime generalized Kronecker delta is $ \delta^{a_1\ldots a_p}_{b_1\ldots b_p} = p!\delta^{a_1}_{[b_1}\cdots \delta^{a_p}_{b_p]} $. We define the projections
\begin{equation}
\bar{\delta}^{a_1\ldots a_p}_{b_1\ldots b_p} = p!\bar{\delta}^{a_1}_{[b_1}\cdots \bar{\delta}^{a_p}_{b_p]},\quad \widetilde{\delta}^{a_1\ldots a_p}_{b_1\ldots b_p} = p!\widetilde{\delta}^{a_1}_{[b_1}\cdots \widetilde{\delta}^{a_p}_{b_p]}, \quad \widehat{\delta}^{a_1\ldots a_p}_{b_1\ldots b_p} = p!\widehat{\delta}^{a_1}_{[b_1}\cdots \widehat{\delta}^{a_p}_{b_p]}
\end{equation}
where $ \bar{\delta}^a_b = \delta^a_b + u^au_b $, $ \widetilde{\delta}^a_b = \widetilde{\delta}^a_b - n^an_b $, and $ \widehat{\delta}^a_b = \delta^a_b + u^au_b - n^an_b $. They satisfy
\begin{equation}
\bar{\delta}^{a_1\ldots a_p}_{b_1\ldots b_p} = -u_{a}u^{b}\delta^{aa_1\ldots a_p}_{bb_1\ldots b_p}, \quad \widetilde{\delta}^{a_1\ldots a_p}_{b_1\ldots b_p} = n_an^b\delta^{aa_1\ldots a_p}_{bb_1\ldots b_p},
\label{gkdconv}
\end{equation}
and
\begin{equation}
\widehat{\delta}^{a_1\ldots a_p}_{b_1\ldots b_p} = n_an^b\bar{\delta}^{aa_1\ldots a_p}_{bb_1\ldots b_p} = -u_au^b\widetilde{\delta}^{aa_1\ldots a_p}_{bb_1\ldots b_p}.
\end{equation}
A tensor $ T $ projected onto $ \Sigma $, $ \Gamma $, or  $ \mathcal{B} $ is indicated as $ \bar{T} $, $ \widetilde{T} $, or $ \widehat{T} $ respectively. The spatial metric is usually expanded in inverse powers of the radial coordinate as $ \gamma_{\alpha\beta} = \delta_{\alpha\beta} + \widetilde{\gamma}_{\alpha\beta} $ where $ \widetilde{\gamma}_{\alpha\beta} $ is not to be confused with the notation $ \widetilde{T} $ for projections onto $ \Gamma $.

\section{Pure Lovelock gravity and ADM mass} \label{puresurfs} \label{purelove} \label{sectadm}

Lovelock gravity is the unique theory constructed out of the Riemann tensor whose equations of motion are second order in the derivatives of the metric. A general Lovelock action is a sum of Lovelock invariants that contain higher powers of the contractions of the Riemann tensor. The simplest examples are Einstein gravity in four dimensions (general relativity) and Gauss-Bonnet gravity in five dimensions. Pure Lovelock gravity in $ D $-dimensions is the theory that contains only a single Lovelock invariant labeled by an integer $ m $. Let $ \mathcal{M} $ be a $ D $-dimensional spacetime. The action is
\begin{equation}
I^\text{bulk}_{(m)} = \frac{1}{16\pi G_\text{N}}\int_{\mathcal{M}}\sqrt{-g}\, c_{(m)} \mathcal{L}_{(m)}
\label{bulkact}
\end{equation}
where the Lovelock invariant
\begin{equation}
\mathcal{L}_{(m)} = \frac{1}{2^m} \delta^{a_1b_1\ldots a_mb_m}_{c_1d_1\ldots c_md_m}R^{c_1d_1}_{a_1b_1}\cdots R^{c_md_m}_{a_mb_m}
\label{bulklagr}
\end{equation}
such that $ \mathcal{L}_{(0)} = 1 $.\footnote{The indices $ a,b = 0,1,\ldots , D-1 $ label the $ D $ spacetime coordinates.} The generalized Kronecker delta (gKd) is defined as
\begin{equation}
\delta^{a_1\ldots a_p}_{b_1\ldots b_p} \equiv \epsilon^{a_1\ldots a_p} \epsilon_{b_1\ldots b_p} = p!\delta^{a_1}_{[b_1}\cdots \delta^{a_p}_{b_p]}
\end{equation}
where $ \epsilon_{a_1\ldots a_p} $ is the Levi-Civita symbol. The parameter $ c_{(m)} $ has length dimension $ 2(m-1) $ to compensate for the additional Riemann tensors in the Lagrangian. From here on out, we will set $ c_{(m)} \slash G_\text{N} = 1 $ which in the case of Einstein gravity coincides with the convention $ G_\text{N} = 1 $.

The Lagrangian \eqref{bulklagr} leads to non-trivial dynamics only in $ D > 2m+1 $ dimensions. In the critical dimension $ D=2m+1 $, the Lagrangian is a total derivative \cite{yale_structure_2011} and for $ D < 2m+1 $ it is identically zero. In this paper, we will be working in dimensions strictly greater than the critical dimension $ D=2m+1 $.

Suppose now that $ \mathcal{M} $ has a timelike boundary $ \Gamma $. The variation of the pure Lovelock action is given by (see Appendix A)
\begin{equation}
\delta I^\text{bulk}_{(m)} = \frac{1}{16\pi}\int_\mathcal{M}\sqrt{-g}\, E_{(m)}^{ab}\delta g_{ab} + \frac{1}{16\pi}\int_{\Gamma} \sqrt{-h}\, \Theta_{(m)}
\label{actvar}
\end{equation}
where the equation of motion tensor
\begin{equation}
E^a_{b(m)}=- \frac{1}{2}\frac{1}{2^{m}}\delta^{aa_1b_1\ldots a_mb_m}_{bc_1d_1\ldots c_md_m}R^{c_1d_1}_{a_1b_1}\cdots R^{c_md_m}_{a_mb_m}
\label{EOM}
\end{equation}
and the boundary term
\begin{equation}
\Theta_{(m)} = -2P^{ab}_{cd(m)}n_{a}\nabla^c\delta g_b^d\,.
\label{lovebound}
\end{equation}
Here $ n^a $ is the spacelike unit normal of $ \Gamma $ and $ \delta g^a_b = g^{ac}\delta g_{cb} = (g^{-1}\delta g)^a_b $. The tensor $ P^{ab}_{cd(m)} $ is the derivative of the Lagrangian with respect to the Riemann tensor
\begin{equation}
P^{ab}_{cd(m)} \equiv \frac{\partial \mathcal{L}_{(m)}}{\partial R^{cd}_{ab}} =\frac{m}{2}\frac{1}{2^{m-1}}\delta^{aba_1b_1\ldots a_{m-1}b_{m-1}}_{cdc_1d_1\ldots c_{m-1}d_{m-1}}R^{c_1d_1}_{a_1b_1}\cdots R^{c_{m-1}d_{m-1}}_{a_{m-1}b_{m-1}}.
\label{lovelockP}
\end{equation}
It is divergence-free in all of its indices $ \nabla_aP^{ab}_{cd(m)} = 0  $, which is the reason why the equations of motion are of second order in the metric.

\subsection{Hamiltonian decomposition}

The ADM mass of pure Lovelock gravity can be obtained by the same method as in Einstein gravity \cite{regge_role_1974}. However, there are subtleties that did not exist in Einstein gravity. These subtleties are absent in asymptotically AdS spacetimes which was the case studied in \cite{kastor_mass_2011}. In that case, it was found that the mass in Lovelock gravity is actually proportional to the Einstein one \cite{kastor_mass_2011}. We will first focus on asymptotically flat spacetimes not studied in \cite{kastor_mass_2011}, and reconsider the simpler asymptotically AdS spacetimes in section \ref{adscase}.

Let us now formulate the theory in the Hamiltonian formalism. To do this, we foliate the spacetime with spacelike constant-$ t $ slices $ \Sigma_t $ that have a timelike unit normal $ u^a $. We are solely interested in static spacetimes so we assume that the shift vector vanishes everywhere. As a result, the metric has the ADM decomposition
\begin{equation}
ds^2 = -N^2dt^2 + \gamma_{\alpha\beta} dx^\alpha dx^\beta
\label{admmetric}
\end{equation} 
where $ N $ is the lapse and $ \gamma_{\alpha\beta} $ is the spatial metric. The canonical coordinate is taken to be the spatial metric, whose velocity $ \dot{\gamma}_{ab} = \pounds_u \gamma_{ab} $ is proportional to the extrinsic curvature of the spacelike slice $ \Sigma_0 \equiv \Sigma $.

The Hamiltonian formulation of Lovelock gravity was laid out first time in \cite{teitelboim_dimensionally_1987} where no timelike boundary was assumed. In the ADM decomposition \eqref{admmetric} of the metric, the bulk action \eqref{bulkact} can be written as \cite{teitelboim_dimensionally_1987}
\begin{equation}
I_{(m)}^\text{bulk} = \frac{1}{16\pi}\int dt\int_{\Sigma_t}\sqrt{\gamma}\, \left[ 2\pi^{ab}_{(m)}\dot{\gamma}_{ab}-N\mathcal{H}_{(m)}\right]
\label{bulkactm}
\end{equation}
where we assumed that there is no timelike boundary. Here $ \pi^{ab}_{(m)} $ is the canonical momentum and
\begin{equation}
\mathcal{H}_{(m)} = -2E_{b(m)}^au_au^b = -\frac{1}{2^{m}}\bar{\delta}^{a_1b_1\ldots a_mb_m}_{c_1d_1\ldots c_md_m}R^{c_1d_1}_{a_1b_1}\cdots R^{c_md_m}_{a_mb_m}.
\label{hambulk}
\end{equation}
where the extra minus sign is due to the contribution of the signature in $ \bar{\delta} $ \eqref{gkdconv}. Here the spacetime Riemann tensors depend on the extrinsic curvature of the spacelike slice $ \Sigma $ through the Gauss-Codazzi equation. The canonical momentum $ \pi^{ab}_{(m)} $ is a complicated polynomial expression in the extrinsic curvature of $ \Sigma $ \cite{teitelboim_dimensionally_1987}. This means that one cannot invert the relation to obtain a unique expression for the extrinsic curvature (velocity) in terms of $ \pi^{ab}_{(m)} $. This is not an issue for deriving an ADM mass, but is a problem for the Hamiltonian formulation as the Hamiltonian is not single valued in the momentum.

From \eqref{bulkactm} the bulk Hamiltonian can be identified to be
\begin{equation}
H^{\text{bulk}}_{(m)} = \frac{1}{16\pi}\int_\Sigma \sqrt{\gamma}\, N \mathcal{H}_{(m)}.
\label{bulkham}
\end{equation}

\subsection{Hamilton's equations and the ADM mass}

We now have the necessary ingredients to derive the ADM mass of pure Lovelock gravity along the lines of \cite{regge_role_1974} where it is defined as the on-shell value of the Hamiltonian.\footnote{See \cite{deser_gravitational_2002,deser_energy_2003,senturk_energy_2012,amsel_wald-like_2013} for related work on conserved charges of higher curvature theories of gravity that extend the results of \cite{abbott_stability_1982} derived in Einstein gravity.} However, \eqref{bulkham} is not the correct Hamiltonian, and has to be modified, as it does not lead to the correct Hamilton's equations which are equivalent to the equations of motion. Let us see how the Hamilton's equations are violated by calculating the variation $ \delta H^{\text{bulk}}_{(m)} $ with respect to the canonical coordinate $ \gamma_{ab} $. To calculate the variation, we expand \eqref{hambulk} using the Gauss-Codazzi equation as
\begin{equation}
\mathcal{H}_{(m)} = -\bar{\mathcal{L}}_{(m)} + \text{extrinsic curvatures of }\Sigma,
\label{Hm}
\end{equation}
where $ \bar{\mathcal{L}}_{(m)} $ is the Lagrangian \eqref{bulklagr} constructed out of the spatial metric $ \gamma_{ab} $. We focus on static spacetimes with a timelike isometry so the terms proportional to the extrinsic curvature vanish.\footnote{The analysis goes through also for stationary spacetimes that have a timelike isometry at infinity. In that case, all the additional boundary terms (ones that do not come from $ \bar{\mathcal{L}}_{(m)} $) are multiplied by at least one factor of extrinsic curvature which vanishes at infinity. Hence the only boundary term that survives comes from $ \bar{\mathcal{L}}_{(m)} $ as in \eqref{spatvar}.} Hence we only keep the first term in \eqref{Hm} and similarly to the spacetime variation of $ \mathcal{L}_{(m)} $ \eqref{actvar}, we get
\begin{equation}
\delta H^{\text{bulk}}_{(m)} =  -\frac{1}{16\pi}\int_\Sigma \sqrt{\gamma}\, N \mathcal{A}^{ab}_{(m)}\delta \gamma_{ab} + \frac{1}{16\pi}\int_\infty \sqrt{\sigma}\, 2\bar{P}^{ab}_{cd(m)}n_aD^c \delta \gamma^d_b
\label{spatvar}
\end{equation}
where the form of $ \mathcal{A}_{ab(m)} $ is not relevant. The bar indicates that the gKd in $ \bar{P}^{ab}_{cd(m)} $ is the spatial one $ \bar{\delta} $. We also assumed asymptotic flatness $ N=1 $ at infinity. The boundary term is interpreted as an integral over a cut-off surface with the cut-off taken to infinity. Its vanishing depends on the asymptotic fall-off conditions of the metric. For the asymptotically flat fall-off conditions specified below, it is non-vanishing and cannot be neglected. The variation \eqref{spatvar} has almost the correct form of a Hamilton's equation
\begin{equation}
-\frac{\delta H^\text{bulk}_{(m)}}{\delta \gamma_{ab}} = \mathcal{A}_{(m)}^{ab} + \text{a boundary term},
\label{hameq}
\end{equation}
but the additional boundary term spoils it. We can define a new Hamiltonian $ H_{(m)} $ that obeys the correct Hamilton's equations by subtracting the boundary term as
\begin{equation}
\delta H_{(m)} = \delta H^\text{bulk}_{(m)} - \frac{1}{16\pi}\int_\infty \sqrt{\sigma}\, 2\bar{P}^{ab}_{cd(m)}n_aD^c \delta \gamma^d_b\,.
\end{equation}
The equation cannot be integrated for $ H_{(m)} $ in general as we see below. Regardless, defining the perturbative ADM mass $ \delta M_{(m)}^\text{ADM} $ as the on-shell value of this Hamiltonian, we get by the diffeomorphism constraint $ H^\text{bulk}_{(m)} = 0 $ that  
\begin{equation}
\delta M_{(m)}^\text{ADM} = -\frac{1}{16\pi}\int_\infty\sqrt{\sigma}\, 2\bar{P}^{ab}_{cd(m)}n_aD^c \delta \widetilde{\gamma}^d_b
\label{asympbound}
\end{equation}
where we have expanded the spatial metric as $ \gamma_{\alpha\beta} = \delta_{\alpha\beta} + \widetilde{\gamma}_{\alpha\beta} $ (the tilde here does not refer to a boundary projection). Equation \eqref{asympbound} is not integrable in general, with the non-integrable piece proportional to the variation of $ \bar{P}^{ab}_{cd(m)} $:
\begin{equation}
\delta M_{(m)}^\text{ADM} = -\frac{1}{16\pi}\delta\int_\infty\sqrt{\sigma}\, 2\bar{P}^{ab}_{cd(m)}n_aD^c \widetilde{\gamma}^d_b - \frac{1}{16\pi}\int_\infty\sqrt{\sigma}\, 2\delta\left( \bar{P}^{ab}_{cd(m)}\right) n_aD^c \widetilde{\gamma}^d_b\,.
\label{integrable}
\end{equation}
For Einstein gravity $ m=1 $, the non-integrable piece vanishes since $ \bar{P}^{ab}_{cd(1)} = (1\slash 2)\bar{\delta}^{ab}_{cd} $ and the mass can be integrated:
\begin{equation}
M_{(1)}^\text{ADM} = \frac{1}{16\pi}\int_\infty\sqrt{\sigma}\, n^a\left( D_b  \widetilde{\gamma}_{a}^b - D_a \widetilde{\gamma} \right).
\label{einsteinadm}
\end{equation}
This is the usual expression for the ADM mass in general relativity \cite{regge_role_1974,hawking_gravitational_1996}.

Obtaining a non-perturbative expression for $ M_{(m)}^\text{ADM} $ in pure Lovelock gravity of order $ m\leq 2 $ cannot be done in general. However, in the presence of a cosmological constant, $ \bar{P}^{ab}_{cd(m)} $ is asymptotically a constant and \eqref{integrable} becomes integrable (see section \ref{adscase}). The formula \eqref{asympbound} is also integrable for asymptotically flat spacetimes that are spherically symmetric, which is seen explicitly in section \ref{expsph}. Regardless of this issue, \eqref{asympbound} is a formula for the perturbation of the ADM mass and it is exactly the quantity that appears in the black hole first law of pure Lovelock black holes.\footnote{By demanding a first law, a formula for black hole entropy in Lovelock gravity was derived in \cite{jacobson_entropy_1993}.}

\subsection{Asymptotic fall-off conditions}

The mass \eqref{asympbound} is finite given a generalization of asymptotic flatness to pure Lovelock gravities. For these theories, the decay of the metric is required to be slower than in Einstein gravity
\begin{equation}
\widetilde{\gamma}_{ab} = \mathcal{O}\left(r^{-\beta} \right) , \quad \beta = \frac{D-(2m+1)}{m},
\label{loveflat}
\end{equation}
because of the decaying Riemann tensors in $ \bar{P}^{ab}_{cd(m)} $. (Note that $ \delta \widetilde{\gamma}_{ab} $ has the same asymptotic behaviour as $ \widetilde{\gamma}_{ab} $ since $ \delta $ only acts on the parameters of the metric.) Indeed given this behaviour, the spacetime Riemann tensor goes as
\begin{equation}
R^{ab}_{cd} = \mathcal{O}\left( r^{-(\beta + 2)}\right) 
\label{riemannbeh}
\end{equation}
so that $ \bar{P}^{ab}_{cd(m)} \sim r^{-(\beta + 2)(m-1)} $. The exponent $ \beta $ has the non-trivial property that
\begin{equation}
-(\beta + 2)(m-1) = (\beta + 1) - (D-2)
\label{betaid}
\end{equation}
so that $ \bar{P}^{ab}_{cd(m)} \sim r^{(\beta + 1) - (D-2)} $. The first factor cancels the contribution coming from the derivative of the metric $ D^c \delta \widetilde{\gamma}^d_b \sim r^{-(\beta + 1)} $ in \eqref{asympbound}, while the second factor cancels the contribution $ r^{D-2} $ coming from the integration measure.

\section{Boundary terms and quasi-local energy} \label{sectquas}

We saw how the bulk Hamiltonian $ H^\text{bulk}_{(m)} $ obtained form the bulk action $ I^\text{bulk}_{(m)} $ does not lead to the correct Hamilton's equations, because an additional boundary term arises. This problem already appears in the Lagrangian formulation as the variational principle is not well defined due to the boundary term in \eqref{actvar}: the Dirichlet variation $ \delta I^\text{bulk}_{(m)} = 0 $, with the boundary metric kept constant, is not equivalent with the equations of motion $ E^a_{b(m)} = 0 $. 

This problem is fixed by adding to the action a generalized Gibbons-Hawking surface term $ I^\text{srf}_{(m)} $ that cancels the boundary terms arising from the variation. The existence of such surface term lies in the fact that the bulk action $ I^\text{bulk}_{(m)} $ is equal to the Euler characteristic in the critical dimension $ D=2m+1 $: it is a topological invariant of a manifold without a boundary. If the manifold also has a boundary, an additional term, the Chern-Simons form, has to be added to create a topological invariant. The Chern-Simons form is then the generalized Gibbons-Hawking term $ I^\text{srf}_{(m)} $. This is because a topological invariant is invariant under metric perturbations that keep the boundary metric fixed. Thus any occurring boundary terms have cancelled. The cancellation is independent of the dimension, providing us with the Gibbons-Hawking term also in higher dimensions $ D > 2m+1 $ \cite{padmanabhan_lanczos-lovelock_2013}.

\subsection{Gibbons-Hawking surface term in the metric formulation}

The first construction of the Gibbons-Hawking term in Lovelock gravity was done by Myers \cite{myers_higher-derivative_1987} in the vielbein formulation, where it is given by the Chern-Simons differential form. Here we will stay in the metric formulation, in which a particularly straightforward derivation of the Gibbons-Hawking term was given in \cite{chakraborty_novel_2017}, which we now review.

First it is shown that the boundary term \eqref{lovebound} can be written as\footnote{We have a different overall sign in \eqref{bndry}, \eqref{var} and \eqref{bterm} compared to \cite{chakraborty_novel_2017}. This choice is consistent with standard Einstein gravity formulas, namely, our conventions are such that \eqref{bterm} and \eqref{boundarystress} reduce to them for $ m=1 $.}
\begin{equation}
\Theta_{(m)} = 2A^{\mu}_{\nu(m)}\delta K^\nu_\mu + B^\mu_{\nu (m)}\delta h^{\nu}_\mu
\label{bndry}
\end{equation}
where $ h_{\mu\nu} $ is the induced metric of the timelike boundary $ \Gamma $ and $ K_{\mu\nu} = (1\slash 2)\pounds_n h_{\mu\nu} $ is the extrinsic curvature of $ \Gamma $.\footnote{The Greek indices $ \mu,\nu,\ldots = 0,1,\ldots, D - 1 $ label the coordinates of the boundary $ \Gamma $.} Here
\begin{equation}
A^{\mu}_{\nu(m)} = -2P^{a\mu}_{b\nu(m)}n_an^b 
\label{Aeom}
\end{equation}
and the expression for $ B^\mu_{\nu (m)} $ can be found in \cite{chakraborty_novel_2017}. There is also a total derivative term that we do not write explicitly.

Including the volume element, \eqref{bndry} can be manipulated to the final form \cite{chakraborty_novel_2017}
\begin{equation}
\sqrt{-h}\,\Theta_{(m)} = -\delta \left( \sqrt{-h}\,B_{(m)} \right) + \tau_{\mu\nu(m)}\delta h^{\mu\nu}.
\label{var}
\end{equation}
Again the total derivative term has been neglected as it can be integrated to the future and past boundaries.\footnote{This total derivative term has recently been incorporated in the covariant phase space formalism \cite{harlow_covariant_2019} where it appears as the exterior derivative $ dC $.} The quantities in \eqref{var} are \cite{chakraborty_novel_2017,olea_mass_2005,miskovic_counterterms_2007,deruelle_einstein-gauss-bonnet_2018}
\begin{equation}
B_{(m)} = 2m\int_0^1du\,\delta^{\mu\mu_1\nu_1\ldots \mu_{m-1}\nu_{m-1}}_{\nu\rho_1\sigma_1\ldots \rho_{m-1}\sigma_{m-1}}K^{\nu}_{\mu}\mathcal{R}^{\rho_1\sigma_1}_{\mu_1\nu_1} \cdots \mathcal{R}^{\rho_{m-1}\sigma_{m-1}}_{\mu_{m-1}\nu_{m-1}}
\label{bterm}
\end{equation}
and
\begin{equation}
\tau^\mu_{\nu(m)} = m\int_0^1du\,\delta^{\mu\rho\mu_1\nu_1\ldots \mu_{m-1}\nu_{m-1}}_{\nu\sigma \rho_1\sigma_1\ldots \rho_{m-1}\sigma_{m-1}}K^{\sigma}_\rho\mathcal{R}^{\rho_1\sigma_1}_{\mu_1\nu_1} \cdots \mathcal{R}^{\rho_{m-1}\sigma_{m-1}}_{\mu_{m-1}\nu_{m-1}},
\end{equation}
with
\begin{equation}
\mathcal{R}^{\rho\sigma}_{\mu\nu} =  \frac{1}{2}\widetilde{R}^{\rho\sigma}_{\mu\nu} - u^2K^{\rho}_{[\mu}K^{\sigma}_{\nu]}.
\label{mathr}
\end{equation}
$ \widetilde{R}^{\rho\sigma}_{\mu\nu} $ is the boundary Riemann tensor (constructed from the boundary metric). The Gibbons-Hawking surface term in pure Lovelock gravity is defined to be
\begin{equation}
I^\text{srf}_{(m)} = \frac{1}{16\pi}\int_{\partial \mathcal{M}}\sqrt{-h}\,B_{(m)}.
\label{gibbonshawking}
\end{equation}
The variation of the total action $ I_{(m)} = I^\text{bulk}_{(m)} + I^{\text{srf}}_{(m)} $, where the bulk action is given in \eqref{bulkact}, becomes
\begin{equation}
\delta I_{(m)} = \frac{1}{16\pi}\int_\mathcal{M}\sqrt{-g}\, E_{(m)}^{ab}\delta g_{ab} + \frac{1}{16\pi}\int_{\Gamma} \sqrt{-h}\, \tau^\mu_{\nu(m)}\delta h^\nu_\mu\,.
\label{totvar}
\end{equation}
Keeping the boundary metric fixed $ \delta h_{\mu\nu} = 0 $, the equations of motion are thus obtained from the variation $ \delta I_{(m)} = 0 $ of the total action. The Gibbons-Hawking surface term has cancelled the first term of \eqref{var}. From \eqref{totvar} we also see that $ \tau_{\mu\nu(m)} $ is proportional to the boundary stress-energy tensor
\begin{equation}
\frac{2}{\sqrt{-h}}\frac{\delta I_{(m)}}{\delta h^{\mu\nu}}\bigg|_\text{on-shell} = \frac{1}{8\pi}\tau_{\mu\nu (m)} .
\label{boundarystress}
\end{equation}

\subsection{Hamiltonian decomposition with a timelike boundary}

Above in the derivation of the ADM mass, we considered the Hamiltonian decomposition when there is no timelike boundary present. This time we will include the timelike boundary $ \Gamma $ in the analysis, and see how the Hamiltonian decomposition is modified by the inclusion of the Gibbons-Hawking term in the action. When there is no boundary, the bulk action has the form \eqref{Hm}. We propose that the decomposition of the total action should then have the form
\begin{equation}
I_{(m)} = \frac{1}{16\pi}\int dt\int_{\Sigma}\sqrt{\gamma}\, \left[ 2\pi^{ab}_{(m)}\dot{\gamma}_{ab}-N\mathcal{H}_{(m)}\right] + \frac{1}{16\pi}\int_\mathcal{B}\sqrt{\sigma}\,N\left[\widehat{B}_{(m)} + \mathcal{K}_{(m)}\right] 
\label{decompbound}
\end{equation}
where we have added two additional boundary terms. $ \mathcal{K}_{(m)} $ contains terms proportional to the extrinsic curvature of the Cauchy slice $ \Sigma $ and thus vanishes when evaluated on static spacetimes. The relevant boundary piece in static spacetimes is $ \widehat{B}_{(m)} $ which is the generalized Gibbons-Hawking surface term \eqref{bterm} of the codimension-two surface $ \mathcal{B} $. Explicitly
\begin{equation}
\widehat{B}_{(m)} = 2m\int_0^1du\,\delta^{ii_1j_1\ldots i_{m-1}j_{m-1}}_{jk_1l_1\ldots k_{m-1}l_{m-1}}\widehat{K}^{j}_{i}\widehat{\mathcal{R}}^{k_1l_1}_{i_1j_1} \cdots \widehat{\mathcal{R}}^{k_{m-1}l_{m-1}}_{i_{m-1}j_{m-1}}
\label{spatbound}
\end{equation}
where
\begin{equation}
\widehat{\mathcal{R}}^{kl}_{ij} =  \frac{1}{2}\widehat{R}^{kl}_{ij} - u^2\widehat{K}^{k}_{[i}\widehat{K}^{l}_{j]}\,.
\end{equation}
Here $ \widehat{R}^{kl}_{ij}  $ is the Riemann tensor and $ \widehat{K}_{ij} $ is the extrinsic curvature of $ \mathcal{B} $.\footnote{The latin indices $ i,j = 1,2,\ldots, D-2 $ label the $ D-2 $ coordinates of $ \mathcal{B} $.} 

For the Lagrangian and Hamiltonian formulations to be consistent, the Hamilton's equations should not contain any boundary terms. We can check this for \eqref{decompbound} by calculating the Hamiltonian (dropping the $ \mathcal{K}_{(m)} $ which is not relevant for static spacetimes):
\begin{equation}
H_{(m)} =  \frac{1}{16\pi}\int_\Sigma \sqrt{\gamma}\, N\mathcal{H}_{(m)} - \frac{1}{16\pi}\int_\mathcal{B}\sqrt{\sigma}\,N\widehat{B}_{(m)}
\label{loveham}
\end{equation}
The first term is the bulk Hamiltonian \eqref{bulkham} encountered before, but now there is an additional boundary term that cancels all the resulting boundary contributions of the variation in static spacetimes. This follows from the fact that the static piece of $ \mathcal{H}_{(m)} $ \eqref{Hm} is the spatial Lagrangian $ \bar{\mathcal{L}}_{(m)} $ (constructed out of the spatial metric) whose variation is cancelled by $ \widehat{B}_{(m)} $. The boundary term $ \widehat{B}_{(m)} $ is the unique term that ensures consistency between the two formulations for static spaces.

\subsection{Quasi-local energy}

Now one could define a quasi-local energy as the on-shell value of the Hamiltonian which would be the integral $ \int_\mathcal{B}\sqrt{\sigma}\,N\widehat{B}_{(m)} $. However, this integral does not have a finite limit when the surface is taken to spatial infinity. Thus the Hamiltonian has to be regulated in some way, while at the same time preserving the form of the Hamilton's equations. This can be done using the so called background subtraction \cite{brown_quasilocal_1993,hawking_gravitational_1996}. The idea is to take a reference spacetime, which in the case of asymptotically flat spacetime is simply Minkowski space, and calculate the corresponding value of the Hamiltonian $ H\lvert_{(0)} $. Then one defines a {\it physical Hamiltonian} $ H^\text{phys}_{(m)} $ with the reference value subtracted:\footnote{In quantum theory such differences are natural. The physical Hamiltonian corresponds to the physical action $ I^\text{phys}_{(m)} = I_{(m)} - I_{(m)}\lvert_{(0)} $ and in the canonical quantization, it is only this difference that has observable consequences since absolute values of the phase are not observable.}
\begin{equation}
H^\text{phys}_{(m)} = H^{\text{bulk}}_{(m)} - \frac{1}{16\pi}\int_\mathcal{B}\sqrt{\sigma}\,N\left( \widehat{B}_{(m)} - \widehat{B}_{(m)}\lvert_{(0)}\right).
\label{phys}
\end{equation}
Here $ \widehat{B}_{(m)}\lvert_{(0)} $ is the boundary term of the surface $ \mathcal{B} $ in flat spacetime and we used the diffeomorphism constraint $ H^\text{bulk}_{(m)}\lvert_{(0)} = 0 $. We also assumed that the lapse and the induced metric of the surface is the same in both spaces. 

The background subtraction requires that the surface $ \mathcal{B} $ can be embedded in flat space in the first place, which is not possible in general. However, when the surface is sufficiently large in an asymptotically flat spacetime, its intrinsic geometry is close to a sphere and an embedding to flat space becomes possible. Regardless, the subtraction regularization satisfies $ \delta H^\text{phys}_{(m)} = \delta H_{(m)} $ so the Hamilton's equations are intact.

Brown-York quasi-local energy is finally defined as the on-shell value of this physical Hamiltonian \cite{brown_quasilocal_1993}:
\begin{equation}
M^{(m)}_{\mathcal{B}} = -\frac{1}{16\pi}\int_\mathcal{B}N\sqrt{\sigma}\,\left( \widehat{B}_{(m)} - \widehat{B}_{(m)}\lvert_{(0)}\right).
\label{lovequas}
\end{equation}
This is a generalization of quasi-local energy to pure Lovelock gravity and contains the familiar Einstein gravity formula as a special case $ m=1 $.

In Einstein gravity there is a second way to arrive at this formula, which utilizes the boundary stress-energy tensor \cite{brown_quasilocal_1993}. This turns out to be true in pure Lovelock gravities as well and it works as follows.\footnote{Brown-York boundary stress-energy tensor is related to the covariant Hamiltonian in an arbitrary diffeomorphism invariant theory of gravity \cite{iyer_comparison_1995,harlow_covariant_2019}.} Given a boundary Killing vector $ \xi^\mu $, the vector $ \tau^{\mu\nu}_{(m)}\xi_\nu $ has a vanishing divergence where the boundary stress-energy tensor $ \tau^{\mu\nu}_{(m)} $ is given in \eqref{boundarystress}. The corresponding conserved quantity on the boundary is an integral over the codimension-2 surface $ \mathcal{B} $:
\begin{equation}
Q[\xi] = \frac{1}{8\pi}\int_{\mathcal{B}}N\sqrt{\sigma}\, u^\mu\tau_{\mu\nu(m)}\xi^\nu.
\end{equation}
In our case, the boundary has a timelike isometry $ \xi^\mu = u^\mu $ (static geometry) and the corresponding charge $ Q[u] $ is interpreted as energy. The energy density is given by\footnote{Note the extra minus sign due to the signature in the definition of the gKd $ \widehat{\delta}^{a_1\ldots a_p}_{b_1\ldots b_p} = -u_au^b\widetilde{\delta}^{aa_1\ldots a_p}_{bb_1\ldots b_p} $.}
\begin{equation}
\varepsilon \equiv u^\mu u^\nu\tau_{\mu\nu(m)} = -\frac{1}{2}\widehat{B}_{(m)}.
\end{equation}
where we used the fact that the spacetime is static so that $ K_{ij} = \widehat{K}_{ij} $. Again $ Q[u] $ diverges in the large surface limit and has to be regulated. The background subtraction prescription yields a quasi-local energy
\begin{equation}
M^{(m)}_{\mathcal{B}} = \frac{1}{8\pi}\int_\mathcal{B}N\sqrt{\sigma}\,(\varepsilon - \varepsilon\lvert_{(0)})
\label{vacdens}
\end{equation}
which agrees with $ \eqref{lovequas} $. The background subtraction is equivalent to defining a renormalized boundary stress-energy tensor $ \tau^\text{ren}_{\mu\nu(m)} = \tau_{\mu\nu(m)} - \tau_{\mu\nu(m)}\lvert_{(0)} $. There are other ways of regularizing the charges like the counterterm method \cite{kraus_gravitational_1999}.

\subsection*{A problem with the large surface limit}

The formula \eqref{lovequas} for quasi-local energy looks fine at first, but there is a problem: the quasi-local energy vanishes in the limit $ \mathcal{B}\rightarrow \infty $ for any solution satisfying the asymptotically flat boundary conditions \eqref{loveflat}:
\begin{equation}
-\frac{1}{16\pi}\int_\infty\sqrt{\sigma}\,\left( \widehat{B}_{(m)} - \widehat{B}_{(m)}\lvert_{(0)}\right) = 0, \quad  m\geq 2.
\end{equation}
To see this, we expand the extrinsic curvature tensor as $ \widehat{K}_{ij} = \widehat{K}_{ij}\lvert_{(0)} \; +\; \delta \widehat{K}_{ij} $ where the perturbation is due to $ \widetilde{\gamma}_{ab} $ in the asymptotic expansion of the metric $ \gamma_{\alpha\beta} = \delta_{\alpha\beta} + \widetilde{\gamma}_{\alpha\beta} $. The variation of $ \widehat{B}_{(m)} $ under $ \widetilde{\gamma}_{\alpha\beta} $ can be obtained from the spatial version of \eqref{var}:
\begin{equation}
-\sqrt{\sigma}\,2\bar{P}^{ab}_{cd(m)}n_aD^c \delta \gamma^d_b = -\delta \left( \sqrt{\sigma}\,\widehat{B}_{(m)} \right) + \bar{\tau}_{ij(m)}\delta \sigma^{ij}.
\label{spatialvar}
\end{equation}
The contribution of the perturbation $ \widetilde{\gamma}_{\alpha\beta} $ to $ \sigma_{ij} $ vanishes asymptotically so we can set $ \delta \sigma_{ij} = 0 $. We get the asymptotic expansion
\begin{equation}
\widehat{B}_{(m)} = \widehat{B}_{(m)}\lvert_{(0)}\; +\; 2\bar{P}^{ab}_{cd(m)}\big|_{(0)}n_aD^c \widetilde{\gamma}^d_b\
\end{equation}
where $ \bar{P}^{ab}_{cd(m)}\big|_{(0)} $ is evaluated on the flat background. Because the Riemann tensor of flat spacetime vanishes, the large surface limit of \eqref{lovequas} vanishes:
\begin{equation}
M_\infty^{(m)} = -\frac{1}{16\pi}\int_\infty\sqrt{\sigma}\,2\bar{P}^{ab}_{cd(m)}\big|_{(0)}n_aD^c \widetilde{\gamma}^d_b = 0.
\label{vanishing}
\end{equation}
Einstein gravity $ m=1 $ is the only theory that avoids this problem, because in that case $ \bar{P}^{ab}_{cd(0)} $ is independent of the spatial metric (and the Riemann tensor). In asymptotically AdS spaces this problem is also avoided, which is the subject of section \ref{adscase}.

\subsection*{Perturbative quasi-local energy and ADM mass}

Regardless of the problem, \eqref{lovequas} defines a weakened form of quasi-local energy: the perturbation of the quasi-local energy $ \delta M_\mathcal{B}^{(m)} $ with respect to the parameters of the metric, approaches the perturbation of the ADM mass $ \delta M^\text{ADM}_{(m)} $. Taking the variation of \eqref{lovequas}, and using \eqref{spatialvar} applied to this variation (while keeping $ \sigma_{ij} $), gives
\begin{equation}
\delta M_\mathcal{B}^{(m)} = -\frac{1}{16\pi}\int_\mathcal{B}\sqrt{\sigma}\,2\bar{P}^{ab}_{cd(m)}n_aD^c \delta \widetilde{\gamma}^d_b
\label{pertmass}
\end{equation}
where we used $ \delta \gamma_{ab} = \delta \widetilde{\gamma}_{ab} $. Clearly taking $ \mathcal{B}\rightarrow \infty $ gives the non-integrable ADM mass $ \delta M^\text{ADM}_{(m)} $ \eqref{asympbound}. This is the pure Lovelock gravity generalization of the Einstein gravity statement saying that Brown-York quasi-local energy approaches ADM mass in the large surface limit.\footnote{Note that the $ \mathcal{B}\rightarrow \infty $ of \eqref{pertmass} does not vanish in contrast to \eqref{vanishing}. The reason is that $ \bar{P}^{ab}_{cd(m)} $ in \eqref{pertmass} is not evaluated on the flat background, but instead interpreted as a limit, like in the ADM mass \eqref{asympbound}.}

We can further simplify \eqref{pertmass} by writing it in terms of intrinsic quantities of $ \mathcal{B} $ and extrinsic curvature. Using \eqref{spatialvar} and the spatial version of \eqref{bndry}, we get (while keeping $ \sigma_{ij} $ fixed)
\begin{equation}
2\bar{P}^{ab}_{cd(m)}n_aD^c \delta \gamma^d_b = -4m\widehat{E}^i_{j(m-1)}\delta \widehat{K}_{i}^j
\label{intrspert}
\end{equation}
where the equation of motion tensor is projected onto $ \mathcal{B} $:
\begin{equation}
\widehat{E}^i_{j(m)} = -\frac{1}{2}\frac{1}{2^m}\delta^{ii_1j_1\ldots i_{m}j_{m}}_{j k_1l_1\ldots k_{m}l_{m}}R^{k_1l_1}_{i_1j_1}\cdots R^{k_ml_m}_{i_mj_m}.
\label{eomhat}
\end{equation}
Substituting to \eqref{pertmass} gives
\begin{equation}
\delta M^{(m)}_{\mathcal{B}} =  \frac{m}{4\pi}\int_\mathcal{B}\sqrt{\sigma}\, \widehat{E}^i_{j(m-1)}\delta \widehat{K}_{i}^j.
\label{loveadmlimit}
\end{equation}
By the limit of \eqref{pertmass}, this formula provides a simply way to calculate the ADM mass of asymptotically flat metrics by taking the limit $ \mathcal{B}\rightarrow \infty $:
\begin{equation}
\delta M_{(m)}^\text{ADM} = \lim_{\mathcal{B}\rightarrow \infty}\frac{m}{4\pi}\int_\mathcal{B}\sqrt{\sigma}\, \widehat{E}^i_{j(m-1)}\delta \widehat{K}_{i}^j.
\label{quaslimit}
\end{equation}
We demonstrate this explicitly in section \ref{expsph}.

\subsection{Chakraborty-Dadhich quasi-local energy} \label{diffquas}

In \cite{chakraborty_brown-york_2015}, a definition of quasi-local energy in pure Lovelock gravity was proposed. For $ m=1 $ it correctly reduces to the Einstein version \cite{hawking_gravitational_1996}, however for $ m\geq 2 $, the regularization does not preserve the form of the Hamilton's equations. Regardless, it defines a type of energy in pure Lovelock gravity whose limit is proportional to the ADM mass for asymptotically flat and spherically symmetric spacetimes. For generic metrics, the limit differs from the ADM mass \eqref{asympbound} as we will now show.\footnote{A schematic proof of the limit for spherically symmetric metrics was presented in \cite{chakraborty_brown-york_2015}. We will prove this explicitly and find the exact prefactor in section \ref{expsph}.}

The regularization presented in \cite{chakraborty_brown-york_2015} involves replacing all the extrinsic curvatures in the boundary term $ \widehat{B}_{(m)} $ \eqref{spatbound} with the vacuum subtracted ones $ \Delta \widehat{K}_{ij} = \widehat{K}_{ij} - \widehat{K}_{ij}\lvert_{(0)} $. Clearly it is a non-linear procedure, which is the reason why the form of the Hamilton's equations is not preserved. Explicitly the definition is\footnote{CD stands for Chakraborty-Dadhich.}
\begin{equation}
M_\mathcal{B}^\text{CD} = -\frac{1}{16\pi}\int_\mathcal{B} \sqrt{\sigma}\, \widehat{B}_{(m)}\lvert_{\widehat{K}_{ij} \rightarrow \Delta \widehat{K}_{ij}}.
\label{altquas}
\end{equation}
To calculate the limit when the surface is large, we need to expand the product in $ \widehat{B}_{(m)} $ \eqref{spatbound}. First we use the Gauss-Codazzi equation to write the intrinsic Riemann tensors $ \widehat{R}^{ij}_{kl} $ in terms of the spacetime tensors $ R^{ij}_{kl} $ and extrinsic curvatures of $ \mathcal{B} $ (the extrinsic curvatures of $ \Sigma $ are zero for static spacetimes). Expanding the product and integrating term by term gives \cite{miskovic_counterterms_2007}
\begin{equation}
\widehat{B}_{(m)} = \frac{m!}{2^{m}}\delta^{ii_1j_1\ldots i_{m-1}j_{m-1}}_{j k_1l_1\ldots k_{m-1}l_{m-1}}\Delta\widehat{K}^{j}_{i} \sum_{s=0}^{m-1}C_{s}^{(m)}R^{k_1l_1}_{i_1j_1}\cdots R^{k_{s}l_{s}}_{i_{s}j_{s}}\Delta\widehat{K}^{k_{s+1}}_{i_{s+1}}\Delta\widehat{K}^{l_{s+1}}_{j_{s+1}}\cdots \Delta\widehat{K}^{k_{m-1}}_{i_{m-1}}\Delta\widehat{K}^{l_{m-1}}_{j_{m-1}}
\label{exp}
\end{equation}
where
\begin{equation}
C_{s}^{(m)} =  \frac{4^{m-s}}{s!(2m-2s-1)!!}.
\label{Cs}
\end{equation}
The $ s=0 $ term in the sum of \eqref{exp} has no Riemann tensors and the $ s=m-1 $ term has no extrinsic curvatures in the product.

A product $ \Delta\widehat{K}_{ij}\Delta\widehat{K}_{kl} \sim r^{-2(\beta + 1)} $ decays faster than a single Riemann tensor $ R^{kl}_{ij} \sim r^{-(\beta + 2)} $.\footnote{See equations \eqref{extcurv} and \eqref{extrdiff}.} Therefore in the limit $ \mathcal{B} \rightarrow \infty $, the leading contribution to \eqref{exp} comes from the term $ s = m-1 $ with a single factor of the extrinsic curvature.\footnote{Note that the non-regularized product $ \widehat{K}_{ij}\widehat{K}_{kl} $ decays at the same rate as a Riemann tensor. Hence it is crucial that all the extrinsic curvatures are replaced by $ \Delta\widehat{K}_{ij} $.} By noting that 
\begin{equation}
C^{(m)}_{m-1} = \frac{4}{(m-1)!},
\end{equation}
the leading asymptotic contribution to \eqref{exp} is given by
\begin{align}
&\frac{4m}{2^{m}}\delta^{ii_1j_1\ldots i_{m-1}j_{m-1}}_{j k_1l_1\ldots k_{m-1}l_{m-1}} R^{k_1l_1}_{i_1j_1}\cdots R^{k_{m-1}l_{m-1}}_{i_{m-1}j_{m-1}}\Delta\widehat{K}^{j}_{i}\\
&= -4m\widehat{E}^i_{j(m-1)}\Delta\widehat{K}^{j}_{i}.
\end{align}
Thus the large surface limit of \eqref{altquas} is
\begin{equation}
M_\infty^\text{CD} = \lim_{\mathcal{B}\rightarrow \infty}\frac{m}{4\pi}\int_\mathcal{B} \sqrt{\sigma}\,\widehat{E}^i_{j(m-1)} \left( \widehat{K}_i^j - \widehat{K}_i^j \lvert_{(0)} \right).
\label{altlimit}
\end{equation}
This is finite for the asymptotic fall-off \eqref{loveflat} and agrees with the integrable part of \eqref{quaslimit}, but does not include the non-integrable part. Therefore it does not agree with the ADM mass exactly. However for spherically symmetric metrics, the limit is proportional to the ADM mass as we see in the next section.

\section{ADM masses of spherically symmetric spacetimes} \label{expsph}

In this section, we explicitly calculate the ADM masses of static spherically symmetric solutions of pure Lovelock gravity that are asymptotically flat. We will be using the simple formula \eqref{quaslimit} defined as a limit. All we need to do is to calculate the necessary quantities for a constant-radius surface and then take the radius to infinity. The results agree with the previous calculations in the literature that used different methods specialized to spherically symmetric spaces. For general asymptotically flat metrics we were only able to define the perturbative form of the ADM mass \eqref{asympbound}, or equivalently \eqref{quaslimit}, but we will see that in the spherically symmetric case, the definition can be integrated to yield the full non-perturbative mass.

\subsection{General static spherically symmetric spacetime}

A static spherically symmetric metric can be written as
\begin{equation}
ds^2 = -[1-G(r)]dt^2 + \frac{1}{1-F(r)}dr^2 + r^2d\Omega_{D-2}^2
\end{equation}
where $ G(r) $ and $ F(r) $ are some functions that go to zero as $ r\rightarrow \infty $ (asymptotic flatness). To calculate the ADM mass, we use the formula \eqref{quaslimit}. We choose $ \mathcal{B} $ to be a constant-$ r $ surface and calculate the limit $ r \rightarrow \infty $. The angular components (components of $ d\Omega_{D-2}^2 $) of the Riemann tensor and of the extrinsic curvature are
\begin{equation}
R^{kl}_{ij} = \frac{F(r)}{r^2}\delta^{kl}_{ij}, \quad \widehat{K}^i_j = \frac{1}{r}\sqrt{1-F(r)}\delta^i_j.
\label{extcurv}
\end{equation}
Using $ F(r) \rightarrow 0 $ as $ r\rightarrow \infty $, we get\footnote{The gKd contractions are calculated using \eqref{oneleft} with $ n=D-2 $.}
\begin{equation}
\widehat{E}^i_{j(m-1)}\delta \widehat{K}^i_j \sim \frac{1}{4}\frac{(D-2)!}{(D-2m-1)!} \frac{F(r)^{m-1}}{r^{2m-1}}\delta F(r), \quad r \rightarrow \infty.
\label{ekprod}
\end{equation}
The angular integral contributes $ \Omega_{D-2}r^{D-2} $ so that \eqref{quaslimit} becomes
\begin{equation}
\delta M^\text{ADM}_{(m)} = \frac{m}{16\pi}\Omega_{D-2}\,a_{(m)}\lim_{r\rightarrow \infty} \left[ r^{m\beta} F(r)^{m-1}\delta F(r)\right], \quad a_{(m)} = \frac{(D-2)!}{(D-2m-1)!}.
\label{bm}
\end{equation}
This can be further integrated to give
\begin{equation}
M^\text{ADM}_{(m)} = \frac{1}{16\pi}\Omega_{D-2}\,a_{(m)}\lim_{r\rightarrow \infty} \left[ r^{m\beta} F(r)^{m}\right].
\label{expmass}
\end{equation}
From this expression it is clear that the mass is finite if $ F(r) \sim r^{-\beta} $ as $ r\rightarrow \infty $, which is the behaviour discussed before \eqref{loveflat}.

\subsection*{Spherically symmetric black hole solution}

\noindent A static asymptotically flat black hole solutions is given by \cite{banados_dimensionally_1994,cai_topological_1999,crisostomo_black_2000,cai_black_2006,kastor_komar_2008}
\begin{equation}
G(r) = F(r) =  \alpha r^{-\beta}.
\end{equation}
Clearly the solution is asymptotically flat in the pure Lovelock sense \eqref{loveflat} and therefore has a finite mass. Plugging into the formula \eqref{expmass} gives
\begin{equation}
M_{(m)}^\text{ADM} = \frac{1}{16\pi}\Omega_{D-2}\,a_{(m)}\alpha^m.
\end{equation}
The proportionality to $ \alpha^m $ agrees with literature \cite{kastor_komar_2008}. For $ m=1 $ we get the familiar mass of a Schwarzschild black hole
\begin{equation}
M_{(1)}^\text{ADM} = \frac{D-2}{16\pi}\Omega_{D-2}\,\alpha.
\end{equation}

\subsection*{Chakraborty-Dadhich quasi-local energy}

\noindent We can also compute the limiting behaviour of the Chakraborty-Dadhich quasi-local energy $ M^\text{CD}_\mathcal{B} $ \eqref{altquas} for a spherically symmetric metric as $ \mathcal{B}\rightarrow \infty $. The large surface limit is given by the formula \eqref{altlimit}. Using \eqref{extcurv} we have
\begin{equation}
\widehat{K}^i_j - \widehat{K}^i_j \big|_{(0)} = \frac{1}{r}\left(\sqrt{1-F(r)} - 1 \right)  \delta^i_j
\label{extrdiff}
\end{equation}
so that
\begin{equation}
\widehat{E}^i_{j(m-1)}\left( \widehat{K}_i^j - \widehat{K}_i^j \big|_{(0)}\right) \sim \frac{1}{4}\frac{(D-2)!}{(D-2m-1)!} \frac{F(r)^{m}}{r^{2m-1}}, \quad r \rightarrow \infty.
\end{equation}
The result is
\begin{equation}
M_\infty^\text{CD} = \frac{m}{16\pi}\Omega_{D-2}\,a_{(m)}\lim_{r\rightarrow \infty}\left[  r^{m\beta}F(r)^{m}\right] .
\end{equation}
This differs from the ADM mass \eqref{expmass} by a factor of $ m $.

\section{Energies in asymptotically AdS spaces} \label{adscase}

Above we focused on asymptotically flat spaces and found that the ADM mass is not integrable in pure Lovelock gravities with $ m\geq 2 $. In addition, only a perturbative definition of quasi-local energy is consistent with the asymptotically flat fall-off \eqref{loveflat} of the metric. In asymptotically AdS geometries these issues disappear, because the asymptotic behaviour of solutions coincides with that of solutions of Einstein gravity.

Pure Lovelock gravity action in the presence of a cosmological constant is
\begin{equation}
I^\text{bulk}_{(m)} = \frac{1}{16\pi}\int_{\mathcal{M}}\sqrt{-g}\, \left(\mathcal{L}_{(m)} - 2\Lambda_{(m)}\right)
\end{equation}
where we have again set $ c_{(m)} \slash G_\text{N} = 1 $. Requiring an AdS vacuum fixes $ \Lambda_{(m)} $. Calculating the equations of motion from this action and substituting the AdS Riemann tensor
\begin{equation}
R^{cd}_{ab}\big|_\text{AdS} = -(1 \slash \ell^2)\delta^{cd}_{ab}
\label{adsriem}
\end{equation}
gives
\begin{equation}
\Lambda_{(m)} = \frac{1}{2}\frac{(D-1)!}{(D-2m-1)!}\left(-\frac{1}{\ell^2} \right)^{m}
\end{equation}
where we used \eqref{eomcontr} to calculate the contractions. 

\subsection{ADM mass in AdS space}

The Hamiltonian decomposition of the action is exactly the same as before (with the addition of the constant $ \Lambda_{(m)} $). When calculating $ \delta H^{\text{bulk}}_{(m)} $, the only relevant difference is that the lapse is not unity at infinity. This means that when transforming the total derivative term to a boundary term, there will be additional contributions proportional to $ D^cN  $ and $ D_aD^cN $. However their contributions will be subleading given the asymptotic fall-off conditions of the metric. Neglecting the vanishing extra contributions, the result is
\begin{equation}
\delta H^{\text{bulk}}_{(m)} =  -\frac{1}{16\pi}\int_\Sigma \sqrt{\gamma}\, N \mathcal{A}_{ab(m)}\delta \gamma^{ab} + \frac{1}{16\pi}\int_\infty \sqrt{\sigma}\,  N 2\bar{P}^{ab}_{cd(m)}n_aD^c\delta\gamma^d_b
\end{equation}
where the form of $ \mathcal{A}_{ab(m)} $ is not relevant. At spatial infinity, the spatial metric goes as $ \gamma_{\alpha\beta} = \gamma_{\alpha\beta}^\text{AdS} + \widetilde{\gamma}_{\alpha\beta} $ so all the Riemann tensors can be replaced by the AdS ones \eqref{adsriem}. Hence the tensor $ \bar{P}^{ab}_{cd(m)} $ is a constant:\footnote{The gKd contractions are calculated using \eqref{twoleft} with $ n=D-1 $.}
\begin{equation}
\bar{P}^{ab}_{cd(m)}\big|_\text{AdS} = b_{(m)}\bar{\delta}^{ab}_{cd}, \quad b_{(m)} = \frac{m}{2}\frac{(D-3)!}{(D-2m-1)!}\left(-\frac{1}{\ell^2} \right)^{m-1}.
\label{adsP}
\end{equation}
The ADM mass \eqref{asympbound} gives
\begin{equation}
\delta M^\text{ADM}_{(m)} = -\frac{b_{(m)}}{8\pi}\int_\infty N\sqrt{\sigma}\, \bar{\delta}^{ab}_{cd}\,n_aD^c\delta\widetilde{\gamma}^d_b
\end{equation}
which is integrable:
\begin{equation}
M^\text{ADM}_{(m)} = \frac{b_{(m)}}{8\pi}\int_\infty N \sqrt{\sigma}\, n^a\left( D_b  \widetilde{\gamma}_{a}^b - D_a \widetilde{\gamma} \right).
\label{adsadm}
\end{equation}
This has the same form as the Einstein ADM mass \eqref{einsteinadm} in flat space for all $ m $ (but with $ N\neq 1 $). Therefore it is finite given the asymptotic behaviour (setting $ m=1 $ in $ \beta $)
\begin{equation}
\widetilde{\gamma}_{ab} = \mathcal{O}\left( \frac{1}{r^{D-3}} \right).
\label{adsasymp}
\end{equation}
This behaviour is identical to solutions of Einstein gravity, which is expected, because in AdS space, the linearized equations of motion of pure Lovelock gravity are equal to the linearized Einstein's equations \cite{chakraborty_1/r_2018}. Thus all the solutions have the same asymptotic behaviour regardless of the value of $ m $.\footnote{Note that one can construct Lovelock theories of gravity that are not asymptotically Einstein in the presence of a cosmological constant. These theories have a degenerate AdS vacuum and the simplest example (with maximal degeneracy) is the Lovelock unique vacuum theory \cite{kastor_black_2006}. See \cite{arenas-henriquez_vacuum_2017,arenas-henriquez_mass_2019,arenas-henriquez_black_2019} for a definition of (conformal) mass in these theories.}

\subsection{Quasi-local energy in AdS space}

\noindent For asymptotically flat metrics, the non-perturbative quasi-local energy \eqref{lovequas} defined using the background subtraction method is problematic as discussed above: its limit for large surfaces is zero. The reason behind this result is that the curvature of the flat background vanishes. On the other hand for an AdS background, the Riemann tensor does not vanish meaning that the problem is avoided and \eqref{lovequas} has a non-zero large surface limit.

Quasi-local energy \eqref{lovequas} of an asymptotically AdS metric is
\begin{equation}
M^{(m)}_\mathcal{B}=-\frac{1}{16\pi}\int_\mathcal{B}N\sqrt{\sigma}\,\left( \widehat{B}_{(m)} - \widehat{B}_{(m)}\lvert_\text{AdS}\right).
\label{loveads}
\end{equation}
Note that the background subtraction is now respect to the AdS background.

The asymptotic correction to $ \widehat{B}_{(m)}\lvert_\text{AdS} $ can be calculated as before when we analyzed quasi-local energy in flat space. Using \eqref{spatialvar} and \eqref{intrspert} the result is
\begin{equation}
\delta \widehat{B}_{(m)} = -4m\widehat{E}^i_{j(m-1)}\big|_{\text{AdS}}\delta \widehat{K}_{i}^j
\end{equation}
where $ \delta $ is the variation under the asymptotic perturbation $ \widetilde{\gamma}_{ab} $ around AdS space. We get
\begin{equation}
M^{(m)}_\infty = \frac{1}{4\pi}\int_\infty N\sqrt{\sigma}\, m\widehat{E}^i_{j(m-1)}\big|_{\text{AdS}}\left( \widehat{K}^j_i - \widehat{K}^j_i\big|_\text{AdS} \right) 
\end{equation}
where the spatially projected equation of motion tensor is given in \eqref{eomhat}. Substituting the AdS Riemann tensor \eqref{adsriem}, it becomes explicitly\footnote{The gKd contractions are calculated using \eqref{oneleft} with $ n=D-2 $.}
\begin{equation}
m\widehat{E}^i_{j(m-1)}\big|_{\text{AdS}} = -b_{(m)}\delta^i_j
\label{eominads}
\end{equation}
with $ b_{(m)} $ defined in \eqref{adsP}. We get 
\begin{equation}
M^{(m)}_\infty = -\frac{b_{(m)}}{4\pi}\int_\infty N\sqrt{\sigma}\, \left( \widehat{K} - \widehat{K}\lvert_\text{AdS} \right) 
\label{ADMads}
\end{equation}
where $ \widehat{K} $ is the trace of the extrinsic curvature. This can be shown to match with the ADM mass \eqref{adsadm} as in \cite{hawking_gravitational_1996}. Hence quasi-local energy defined using the background subtraction method has the correct ADM mass limit in asymptotically AdS spacetimes.

Finally, we mention that the Chakraborty-Dadhich quasi-local energy \eqref{altquas} also has the correct ADM mass limit \eqref{ADMads} in asymptotically AdS spaces. This follows by plugging \eqref{eominads} into the large surface limit \eqref{altlimit}.


\section{Summary and discussion} \label{summary}

In this paper, we studied how the Einstein gravity derivations of ADM mass \cite{regge_role_1974} and Brown-York quasi-local energy \cite{brown_quasilocal_1993} generalize to pure Lovelock gravity. We focused on asymptotically flat geometries and found that the ADM mass is not integrable in general. We regularized the quasi-local energy using the background subtraction prescription and found that only the perturbative definition is well defined: the limit of the non-perturbative energy for large surfaces vanishes. Then we proved that the perturbative version has the correct ADM mass limit for large surfaces. In asymptotically AdS spaces, there is no problem and the (non-perturbative) quasi-local energy is proportional to the standard Brown-York energy of Einstein gravity for all values of $ m $.

The vanishing of the large surface limit of the non-perturbative quasi-local energy is related to the non-integrability of the ADM mass: only perturbative energies appear to be well defined in pure Lovelock gravities. This is good enough from the perspective of the black hole first law and the ADM mass used in this paper should appear in the first law of pure Lovelock black holes \cite{jacobson_entropy_1993}.

The problem might also be related to the background subtraction prescription itself which is unsatisfactory for other reasons. First, it requires an ad hoc auxiliary spacetime whose geometry might depend on the situation at hand. Second, the embedding of the surface into the auxiliary spacetime is not possible in general. A more viable method of renormalization for asymptotically AdS spaces is provided by the counterterm method \cite{kraus_gravitational_1999} developed in Einstein gravity. Counterterms have been calculated and applied to asymptotically AdS solutions of Lovelock gravity \cite{dehghani_counterterm_2006,miskovic_counterterms_2007,yale_simple_2011}. However, the application to asymptotically flat spacetimes is less explored \cite{astefanesei_note_2007} especially in Lovelock gravity. It would be interesting to apply the counterterm method to asymptotically flat geometries in pure Lovelock gravity.

\section*{Acknowledgements}

The author thanks Esko Keski-Vakkuri for useful comments and discussions. This work is supported in part by the Academy of Finland grant no 
1297472.

\appendix

\section{Generalized Kronecker delta (gKd) identities}

In this Appendix, the Euclidean Latin indices $ i,j,k,l $ take values over a set of size $ n $. The generalized Kronecker delta (gKd) is defined as
\begin{equation}
\delta^{i_1\ldots i_p}_{j_1\ldots j_p} \equiv \epsilon^{i_1\ldots i_p} \epsilon_{j_1\ldots j_p} = p!\delta^{i_1}_{[j_1}\cdots \delta^{i_p}_{j_p]}.
\end{equation}
and $ \epsilon_{12\ldots p}=1 $. Contraction of a single pair of indices:
\begin{equation}
\delta^{i_1\ldots i_pi_{p+1}}_{j_1\ldots j_pj_{p+1}} = (n-p)\delta^{i_1\ldots i_p}_{j_1\ldots j_p}.
\label{contraction}
\end{equation}
Contraction of multiple pairs of indices:
\begin{equation}
\delta^{i_1\ldots i_p i_{p+1}\ldots i_q}_{j_1\ldots j_p i_{p+1}\ldots i_q} = \frac{(n-p)!}{(n-q)!}\delta^{i_1\ldots i_p}_{j_1\ldots j_p}.
\label{kroncontr}
\end{equation}
Contraction of a gKd of size $ p $ with another of size 2:
\begin{equation}
\delta^{i_1\ldots i_pi_{p+1}i_{p+2}}_{j_1\ldots j_pj_{p+1}j_{p+2}}\delta_{i_{p+1}i_{p+2}}^{j_{p+1}j_{p+2}} = 2!\delta^{i_1\ldots i_pi_{p+1}i_{p+2}}_{j_1\ldots j_pi_{p+1}i_{p+2}}= 2(n-p)(n-p-1)\delta^{i_1\ldots i_p}_{j_1\ldots j_p}.
\label{2contr}
\end{equation}
Applying \eqref{2contr} multiple times and using \eqref{kroncontr} with $ p=1 $ and $ q=2m+1 $ we get
\begin{equation}
\delta^{ii_1j_1\ldots i_{m}j_{m}}_{jk_1l_1\ldots k_{m}l_{m}}\delta^{i_1j_1}_{k_1l_1}\cdots \delta^{i_{m}j_{m}}_{k_{m}l_{m}} = \frac{2^{m}(n-1)!}{(n-2m-1)!}\delta^{i}_{j}.
\label{eomcontr}
\end{equation}
Similarly with $ p=1 $ and $ q = 2m-1 $ we get
\begin{equation}
\delta^{ii_2j_2\ldots i_{m}j_{m}}_{jk_2l_2\ldots k_{m}l_{m}}\delta^{i_2j_2}_{k_2l_2}\cdots \delta^{i_{m}j_{m}}_{k_{m}l_{m}} =  \frac{2^{m-1}(n-1)!}{(n-2m+1)!}\delta^{i}_{j}.
\label{oneleft}
\end{equation}
Finally with $ p=2 $ and $ q = 2m $ we get
\begin{equation}
\delta^{iji_2j_2\ldots i_mj_m}_{klk_2l_2\ldots k_ml_m} \delta^{i_2j_2}_{k_2l_2}\cdots \delta^{i_mj_m}_{k_ml_m} =  \frac{2^{m-1}(n-2)!}{(n-2m)!}\delta^{ij}_{kl}.
\label{twoleft}
\end{equation}

\section{Variation of the pure Lovelock action}

The pure Lovelock Lagrangian is given by
\begin{equation}
\mathcal{L}_{(m)} = \frac{1}{2^m} \delta^{a_1b_1\ldots a_mb_m}_{c_1d_1\ldots c_md_m}R^{c_1d_1}_{a_1b_1}\cdots R^{c_md_m}_{a_mb_m}.
\end{equation}
Let us calculate its metric variation
\begin{equation}
\delta I^\text{bulk}_{(m)} = \frac{1}{16\pi}\int_\mathcal{M}\sqrt{-g}\, E^{ab}_{(m)}\delta g_{ab} + \frac{1}{16\pi}\int_{\partial \mathcal{M}} \sqrt{-h}\, \Theta_{(m)}.
\end{equation}
We have
\begin{equation}
\delta\left( \sqrt{-g}\,\mathcal{L}_{(m)}\right)  = \frac{1}{2^m}\sqrt{-g}\left(-\frac{1}{2}\delta^a_b\delta^{a_1b_1\ldots a_mb_m}_{c_1d_1\ldots c_md_m}\delta g^{b}_aR^{c_1d_1}_{a_1b_1} + m \delta^{a_1b_1\ldots a_mb_m}_{c_1d_1\ldots c_md_m}\delta R^{c_1d_1}_{a_1b_1}\right)R^{c_2d_2}_{a_2b_2}\cdots R^{c_md_m}_{a_mb_m}.
\end{equation}
The first term comes from acting on $ \sqrt{-g} $ and the remaining terms result from acting on the $ m $ Riemann tensors. Now apply the formula $ \delta R^{cd}_{ab} = -2\nabla_{[a}\nabla^{[c}\delta g_{b]}^{d]} + R^{e[c}_{ab}\delta g_{e}^{d]} $, where $ \delta g_{a}^{b} = g^{bc}\delta g_{ac} $. This gives
\begin{align}
\delta\left( \sqrt{-g}\,\mathcal{L}_{(m)}\right) = -\sqrt{-g}\frac{1}{2^m}&\left(\frac{1}{2}\delta^a_b\delta^{a_1b_1\ldots a_mb_m}_{c_1d_1\ldots c_md_m}R^{c_1d_1}_{a_1b_1}\delta g^b_a - m \delta^{a_1b_1\ldots a_mb_m}_{c_1d_1\ldots c_md_m}R^{c_1a}_{a_1b_1}\delta g_{a}^{d_1}\right)R^{c_2d_2}_{a_2b_2}\cdots R^{c_md_m}_{a_mb_m} \nonumber\\
& -\frac{2m}{2^{m}}\sqrt{-g}\,\delta^{a_1b_1\ldots a_mb_m}_{c_1d_1\ldots c_md_m}\left( \nabla_{a_1}\nabla^{c_1}\delta g_{b_1}^{d_1}\right) R^{c_2d_2}_{a_2b_2}\cdots R^{c_md_m}_{a_mb_m} 
\label{twolines} 
\end{align}
where we have removed all the anti-symmetrizations based on the anti-symmetricity of the generalized Kronecker delta. Focus on the first line of \eqref{twolines} for the time being. It can be written as
\begin{equation}
-\sqrt{-g}\,\delta g^b_a\frac{1}{2^m}\left(\frac{1}{2}\delta^a_b\delta^{a_1b_1\ldots a_mb_m}_{c_1d_1\ldots c_md_m}R^{c_1d_1}_{a_1b_1} - m\delta^{d_1}_b \delta^{a_1b_1\ldots a_mb_m}_{c_1d_1\ldots c_md_m}R^{c_1a}_{a_1b_1}\right)R^{c_2d_2}_{a_2b_2}\cdots R^{c_md_m}_{a_mb_m}.
\label{firsterm}
\end{equation}
In terms of the tensor
\begin{equation}
P^{ab}_{cd(m)} \equiv \frac{\partial \mathcal{L}_{(m)}}{\partial R^{cd}_{ab}} = \frac{m}{2^{m}}\delta^{aba_2b_2\ldots a_{m}b_{m}}_{cdc_2d_2\ldots c_{m}d_{m}}R^{c_2d_2}_{a_2b_2}\cdots R^{c_md_m}_{a_mb_m},
\end{equation}
\eqref{firsterm} is
\begin{equation}
-\sqrt{-g}\,\delta g^b_a\left( \frac{1}{2}\delta^a_b\mathcal{L}_{(m)} - P^{a_1b_1}_{bc_1(m)}R^{ac_1}_{a_1b_1}\right).
\label{first}
\end{equation}
Focus now on the second line of \eqref{twolines}. It can be written as
\begin{equation}
-\sqrt{-g}\,2P^{ab}_{cd(m)} \nabla_{a}\nabla^{c}\delta g_{b}^{d} = -\nabla_a\left(2P^{ab}_{cd(m)} \nabla^{c}\delta g_{b}^{d} \right) + 2\nabla_aP^{ab}_{cd(m)}\nabla^{c}\delta g_{b}^{d}\,.
\label{second}
\end{equation}
The divergence of $ P^{ab}_{cd(m)} $ vanishes by the Bianchi identity of the Riemann tensor:
\begin{equation}
\nabla_aP^{ab}_{cd(m)} = \frac{m(m-1)}{2^{m}}\delta^{aba_2b_2\ldots a_mb_m}_{cdc_2d_2\ldots c_md_m} \left( \nabla_{[a_1}R^{c_2d_2}_{a_2b_2]}\right) \cdots R^{c_md_m}_{a_mb_m} = 0
\end{equation}
which is the reason why the equations of motion of pure Lovelock gravity are second order.

Combining \eqref{first} and \eqref{second} gives 
\begin{equation}
\delta\left( \sqrt{-g}\,\mathcal{L}_{(m)}\right) = \frac{1}{16\pi}\int_\mathcal{M}\sqrt{-g}\,\delta g^b_a\left(P^{a_1b_1}_{bc_1(m)}R^{ac_1}_{a_1b_1}- \frac{1}{2}\delta^a_b\mathcal{L}_{(m)}\right) - \frac{1}{16\pi}\int_{\partial \mathcal{M}} \sqrt{-h}\,2P^{ab}_{cd(m)}n_{a}\nabla^c\delta g_b^d.
\end{equation}
From this we can identify the equation of motion tensor
\begin{equation}
E^a_{b(m)} = \mathcal{R}^a_{b(m)} - \frac{1}{2}\delta^a_b\mathcal{L}_{(m)},
\end{equation}
where $ \mathcal{R}^a_{b(m)} \equiv P^{a_1b_1}_{bc_1(m)}R^{ac_1}_{a_1b_1} $, and the boundary contribution
\begin{equation}
\Theta_{(m)} = -2P^{ab}_{cd(m)}n_{a}\nabla^c\delta g_b^d.
\end{equation}
The equation of motion tensor can also be written in another form. From \eqref{firsterm} it follows that
\begin{equation}
E^a_{b(m)} = -\frac{1}{2}\frac{1}{2^m}\left(\delta^a_b\delta^{a_1b_1\ldots a_mb_m}_{c_1d_1\ldots c_md_m} - 2m\delta^{a}_{d_1} \delta^{a_1b_1\ldots a_mb_m}_{c_1b\ldots c_md_m}\right)R^{c_1d_1}_{a_1b_1}\cdots R^{c_md_m}_{a_mb_m}.
\label{newform}
\end{equation}
Now consider the identity
\begin{equation}
\delta^{a_1\ldots a_n}_{b_1\ldots b_n} = \sum_{k=1}^n(-1)^{k+1}\delta^{a_k}_{b_1}\delta^{a_2\ldots a_1\ldots a_n}_{b_2\ldots b_k\ldots b_n}.
\label{kroneckeridentity}
\end{equation}
When applied to \eqref{newform} and using the symmetries of $ R^{c_1d_1}_{a_1b_1}\cdots R^{c_md_m}_{a_mb_m} $, we get
\begin{equation}
E^a_{b(m)} = - \frac{1}{2}\frac{1}{2^{m}}\delta^{aa_1b_1\ldots a_mb_m}_{bc_1d_1\ldots c_md_m}R^{c_1d_1}_{a_1b_1}\cdots R^{c_md_m}_{a_mb_m}.
\end{equation}

\bibliographystyle{JHEP}
\bibliography{quasilocal}

\end{document}